
\magnification 1200
\hoffset=1.6 true cm
\voffset=0.3 true cm
\hsize=14.6 true cm
\vsize=21.4 true cm
\baselineskip=.7 true cm plus .002 true cm minus .001 true cm
\newcount\secno \global\secno=0
\newcount\subsecno \global\subsecno=0 \newcount\numec
\global\numec=0 \newcount\pageflag \global\pageflag=0
\openout0=salida
\nopagenumbers \headline={\ifnum\pageflag=1
\hfil\folio \else \global\pageflag=1\hfil\fi}
\def\romanmayusc#1{\uppercase\expandafter{\romannumeral#1}}
\def\capitulo#1 #2\par{\vfill\eject\vglue 1 cm \centerline{\bf
CAP\'ITULO \romanmayusc#1} \centerline{\bf\uppercase{#2}}\vskip
1.5cm\noindent
\write0{#1}\write0{#2}\write0{\the\pageno}
\global\secno=0\global\pageflag=0}
\def\seccion#1\par{\advance\secno by 1 \vskip 0.7cm \leftline{\bf
\the\secno.\enspace#1}\vskip 0.3cm \noindent
\write0{\the\secno}\write0{#1}\write0{\the\pageno}\global\subsecno=0}
\def\subsec#1\par{\advance\subsecno by 1 \vskip 0.5cm
\leftline{\the\secno.\the\subsecno\enspace{\sl#1}}\noindent
\write0{\the\secno.\the\subsecno}\write0{#1}\write0{\the\pageno}}

\def\ecu#1{\advance\numec by 1
\write1{#1}\write1{the\numec}\eqno(\the\numec)}
   \def\dalam{\hbox{\vrule
height0.2cm width0.1mm depth0mm  \vrule height 0.1mm width0.2cm
depth0mm \vrule height0.2cm width0.1mm depth0mm
\kern-0.2cm\raise0.2cm\hbox{\vrule height 0.1mm width0.2cm
depth0mm}}}   

\def\rz {\sqrt {-g}} \def\pr{{2\alpha \Phi\over \sqrt {-g}}}
\def\pp{{2\alpha\Phi'\over \sqrt {-g}}} \def\pd{{2\alpha \dot
\Phi\over\sqrt {-g}  }} \def\pt{{2\alpha\Phi\over (-g)^{-3/ 2} } }
\def\pdt{{2\alpha \dot \Phi\over (-g)^{3/ 2}  }}
\def\ppt{{2\alpha\Phi'\over (-g)^{3/ 2}}} \def\A{(n^in^j)-\delta^{ij}}
\def\B{(n^ix')} \def\C{(n^iu)} \def\pn{\pi_in^i -\pr (x"n^i)}
\def\pu{\pi u} \def\px{\pi x'}
 \def\ho {pu + \rz +
\pr(u'n^i)^2} \def\hu{\pi u'+px'} 
\vglue 2cm \centerline{\bf HIGHER ORDER ACTION FOR THE INTERACTION}
\centerline{\bf OF THE STRING WITH THE DILATON } \vglue 0.5cm
\centerline{M. A. Lled\'o\footnote*{e-mail: mlledo@dino.conicit.ve}
and A. Restuccia} \centerline{\it {Departamento de F\'{\i}sica.
Universidad Sim\'on Bol\'{\i}var.}} \centerline{\it{Apartado 89000,
Caracas 1080-A. Venezuela.}} \vskip 2cm \centerline{Abstract} The
theory of the string in interaction with a dilaton background field is
analyzed. In the action considered, the metric in the world sheet of
the string is the induced metric, and the theory presents second
order time derivatives. The canonical formalism is developed and it
is showed that first and second class constraints appear. The degrees
of freedoom are the same than for the free bosonic string. The light
cone gauge is used to reduce to the physical modes and to compute the
physical hamiltonian. \vfill \eject

\seccion Introduction.

We consider the theory of the bosonic string in interaction with an
scalar field, the dilaton. Much work has been done on the problem of
constructing string theories on general background fields [1]. The
theory describing the interaction of the bosonic string with the
metric, the antisymmetric field and the dilaton, via the Polyakov
approach, was recently studied in the excellent paper by Buchbinder,
Fradkin, Lyakhovich y Pershin [2]. The theory they consider is given
by the action   $$ S=-\int d^2\xi \rz\bigl \{{1\over
2}g^{ab}\partial_ax^\mu \partial_bx^\nu G_{\mu \nu}(x) +{1\over
2}\epsilon^{ab}\partial_ax^\mu \partial_bx^\nu A_{\mu \nu}(x) +^{(2)}R
\Phi(x)\bigr \}.\eqno(1)  $$ $G_{\mu \nu}(x)$ is the D-dimensional
metric,  $A_{\mu \nu}(x)$ is the antisymmetric field and $\Phi(x)$ is
the  dilaton. $g_{ab}(\xi)$ is the metric of the world-sheet of the
string, and it is considered here as a variable, independent of the
embeding $x^\mu (\xi)$. It is a Polyakov-type action.  $^{(2)}R $ is
the  curvature of the two dimensional submanifold,  associated to the
metric $g_{ab}$. As it is well known, this theory can be consistently
quantized provided the external fields satisfy certain restrictions.
If the only background field is  $G_{\mu \nu}(x)$, the theory is
consistent in $D=26$ and the metric satisfies up to linear order in
the curvature the Einstein equations. Nevertheless if the dilaton is
different from zero (different from constant, indeed) the critical
dimension is
 $D=25$ and the  Einstein equations are modified. The dilaton, as
expected,  changes notably the classical and quantum behaviour of the
system. On one hand the field equations of action (1)  imply that the
metric of the world sheet of the string is the induced metric only
if  $\Phi(x)=\hbox{ctt}$. (In two dimensions the last term in (1) is
the Euler characteristic when  $\Phi(x)=\hbox{ctt}$, so it becomes
irrelevant to the field equations). So the presence of the dilaton
changes this geometrical interpretation of the action. On the other
hand, the  degrees of freedoom of the theory are not the same in both
situations. If $\Phi(x)\ne\hbox{ctt}$, the degrees of freedoom are
$D-1$ (the space-time has dimension $D$); if $\Phi(x)=\hbox{ctt}$ the
degrees of freedoom are $D-2$, as in the free bosonic string. This is
because the term proportional to the curvature breaks the
 invariance of the action  under the rescaling of the metric, unless
it is an irrelevant, total derivative. So it appears another degree
of freedoom, and the limit $\Phi(x)\to \hbox{ctt}$ is not smooth. The
free bosonic string is not a good starting point in order to make a
perturbative treatment of the background fields.  In Ref. [2] the
problem is solved considering a string in interaction with a non
trivial dilaton as the base for the perturbative treatment.

The theory we are considering has a similar action, but now the metric
 $g_{ab}(\xi)$ is not an independent variable, but it corresponds to
the induced metric on the two dimensional surface,  $$
 g_{ab}(\xi) = \partial_ax^\mu\partial_bx^\nu G_{\mu \nu}(x),\eqno(2)
$$ and the geometrical interpretation is guaranteed. It is a
Nambu-Goto type action.  Obviously both actions are not equivalent.
The last one is more complicated, since the term containing the
dilaton has higher derivatives. Indeed we will restrict ourselves to
the case $G_{\mu \nu}(x)=\eta_{\mu \nu}$ and $A _{\mu \nu}(x)=0$, it
is, we will retain the dilaton as the only non trivial background
field. This interaction is complex enough, and we expect a better
understanding of the modifications that the dilaton produces compared
with the free string.

In Section 2. we describe the canonical formalism for higher
derivatives. In Section 3. we apply it to the string in interaction
with the dilaton, obtaining the primary constraints. In Section 4. we
compute the secondary constraints. In Section 5. the first class
constraints are covariantly separated from the second class
constraints and the degrees of freedoom of the theory are computed.
In Section 6. we proceed to fix the light cone gauge and to compute
the Hamiltonian, comparing with other approaches. By using the light
cone gauge we do not need to impose any restriction on the background
fields. It is known that starting from the lagrangean approach of a
string in a background, the light cone gauge is an admissible gauge
provided the background is restricted. In particular $G_{\mu \nu}$
must be a pp-wave. However, in the phase space approach we follow in
this paper, no conditions on Killing vectors of the background is
required. The problem is that the only dependence on the transverse
momentum becomes non quadratic because the dependence of the
background with the coordinate $x^-$. This feature does not allow the
functional integration of the transverse momentum to recover the
lagrangean approach for arbitrary  background. However, this
condition on the background is not a requirement of the canonical
formulation. Finally, in Section 7. we establish our conlusions.

\seccion Higher derivatives.

We use the  generalization of the canonical formalism to higher
derivatives proposed in Ref. [3][4]. For clearness, we briefly resum
here the resut for a Lagrangean depending on second order time
derivatives.

We consider a physical system on an $n$-dimmensional configuration
space. Let $L(q,\dot q, \ddot q)$ be the Lagrangean of the system
depending on the coordinates $(q^1,\dots , q^n)$ and their time
derivatives of order one and two. The Euler Lagrange equations are
obtained applying the Hamilton principle to the functional action, $$
S(q)= \int_{t_i}^{t_f} L(q,\dot q, \ddot q)dt. \eqno(3) $$ This is
equivalent to extremizing the constrained functional, $$ R(q, u, v)=
\int_{t_i}^{t_f} L(q,u,v)dt, \eqno(4) $$ subject to $$ \dot q=u
,\quad \dot u=v. \eqno(5) $$ The constraints are regular, so we can
apply the Lagrange theorem and consider the unconstrained functional
$$
 \int_{t_i}^{t_f} [L(q,u,v) -p(u-\dot q)-\pi (v-\dot u)]dt, \eqno(6)
$$ where we have introduced Lagrange multipliers $p$ and $\pi$. The
canonical moments of the coordinates $(q,u)$ are the corresponding
Lagrange multipliers $(p, \pi)$. The Hamiltonian can be read off from
(6) as a function of two sets of canonically conjugated variables
$(q,p)$ and $(u,\pi)$ and a set of non canonical ones $v$, $$
H(q,u,p,\pi, v)= pu+\pi v -L(q,u,v). \eqno(7) $$ Doing independent
variations of all the variables, one obtains canonical equations of
motion for the canonical coordinates, and in addition one obtains the
set of equations $$ {\partial H\over \partial v}=0.\eqno(8) $$ If the
Lagrangean is singular, the Hessian $$ {\partial^2 H\over \partial
v^i\partial v^j}=-{\partial^2 L\over \partial v^i\partial v^j}\eqno(9)
$$ has rank $r<n$, and some of the equations (8) are primary
constraints. The Dirac procedure to compute the complete set of
constraints follows as usual, if constraints are regular. The
components of $v$ which cannot be calculated play the same role as
the Lagrange multipliers associated to first class constraints.

In the next section we apply the formalism to the string in
interaction with de dilaton.

\seccion Canonical action.

We denote $\xi^0=\tau$ and $\xi^1=\sigma$. A dot means a derivative
with respect to $\tau$ and a prime a derivative with respect to
$\sigma$. $D$ is the dimension of space-time, whose metric is flat,
with signature $(1,-1,....,-1)$. We assume the conditions
$g_{00}=\dot x^2 >0$ y $g_{11}= x'^2 <0$ hold. We denote $^{(2)}R$
simply by $R$, since there is not possibility of confussion.

The lagrangean action is  $$ S=-\int d^2\xi \rz(1 + \alpha\Phi
R).\eqno(10) $$ If the space-time metric is flat, the curvature can
be expressed in terms of the second fundamental form of the surface
as  $$ R=s^{ia}_as^{ib}_b-s^{ia}_bs^{ib}_a.\eqno(11) $$ where
$s^i_{ab}\quad i=1, \dots D-2$ are the components of the second
fundamental form in an orthonormal base of the $D-2$ normal vectors to
the surface, $n^i_\mu$.

The orthonormal vectors satisfy  $$ \eqalign { &\A =0 \cr &\B =0\cr
&(n^i\dot x)=0.\cr}\eqno(12) $$ and the second fundamental form is $$
s^i_{ab}=D_aD_bx^{\mu}n^i_\mu=x^{\mu}_{,a,b}n^i_\mu .\eqno(13) $$ We
have different expressions for the curvature. First, we can express
it in terms of the covariant derivatives $D_a$, independently of the
normal vectors, $$ R=g^{ab}g^{cd}D_aD_bx^\mu D_cD_dx_\mu
-g^{ab}g^{cd}D_aD_cx^\mu D_bD_dx_\mu ,\eqno(14)  $$ or in terms of
them, $$ R={2\over g}(s^i_{00}s^i_{11}-(s^i_{01})^2)={2\over g}\bigl (
(x^\mu_{,0,0}n^i_\mu)
(x^\nu_{,1,1}n^i_\nu)-(x^\nu_{,0,1}n^i_{\nu})^2\bigr ).\eqno(15)  $$
Both expressions are equivalent. The normal vectors $n^{i\mu}$ can be
considered as independent variables only if constraints (12) are
introduced in the action with Lagrange multipliers $$
 \eqalign {-&\int d^2\xi\{\rz -\pr [(\ddot xn^i)(x'n^i)-(\dot
x'n^i)^2]+\cr +&\lambda_{ij}\bigl (\A \bigr )-\mu_i\B -\nu_i (n^i\dot
x)\},\cr}\eqno(16) $$ where $\lambda_{ij},\;\mu_i, \eta_i$ are the
Lagrange multipliers associated with the constraints defining the new
variables as an orthonormal system of normal vectors to the surface.
This considerably simplifies the problem.

Let us compute the Euler-Lagrange equations. When varying the  $n^i$
one obtains relations which allow to compute the Lagrange
multipliers. If these relations are introduced in the equation which
results from varying $x^\mu$, one obtains $$
 \eqalign {&\bigl [{(\dot x x')\dot x^\mu \over \rz}-{\dot x^2
x'^\mu\over \rz}\bigr]'+
 \bigl [{(\dot x x')x'^\mu \over \rz }-{ x'^2 \dot x^\mu\over
\rz}\bigr]^.  +{2\alpha R\partial^\mu\Phi\over \rz } +\cr &+ \bigl
[\pp D_0\dot x^\mu-\pd D_1\dot x^\mu\bigr ]'+ \bigl [\pd
D_1x'^\mu-\pp D_0 x'^\mu\bigr ]^.=0.\cr} \eqno(17)  $$ The auxiliary
variables $n^{i\mu}$ are eliminated. The two first terms in (17) are
the field equations of the free bosonic string. The remaining terms
depend on  $\partial _\mu\Phi$, so, if the dilaton is a constant, the
theory is equivalent to the free  bosonic string.

The canonical analysis follows as in the previous section. We
introduce new variables $u^\mu, v^\mu$, $p_\mu,\pi_\mu,
m_{i\mu},\gamma^{i\mu}$. The canonical action is  $$ S=\int
d^2\xi\bigl [p\dot x+ \pi\dot u +m_{i\mu}\dot n^{i\mu}-{\cal
H}(x,u,n,\gamma,p,\pi,m)\bigr],\eqno(18) $$ where $$  \eqalign
{&{\cal H}(x,u,n,\gamma,p,\pi,m)= {\cal H}_0+ v^\mu(\pi_\mu -\pr
(x"n^i)n^i_\mu)+
 \gamma^{i\mu}m_{i\mu} +\cr&+ \lambda_{ij}\bigl (\A \bigr )-\mu_i\B
-\nu_i (n^i\dot x),\cr}\eqno(19) $$ and $$ {\cal H}_0=\ho .\eqno(20)
$$ The variables $(v^\mu,\gamma^{i\mu}, \lambda_{ij},\mu_i,\nu_i)$
act as multipliers. From here, we can read the primary constraints of
the theory.  For the following analysis, it is convenient to consider
the decomposition of the Lagrange multiplier $v^\mu$, in terms of its
normal and tangential components, $$
 v^\mu=\omega^in^{i\mu} +\Lambda_1 u^\mu + \Lambda_2x'^\mu .\eqno(21)
$$ In such way,   $$ v^\mu(\pi_\mu -\pr (x"n^i)n^i_\mu)=\omega^i(\pn)
+\Lambda_1\pu +\Lambda_2\px . \eqno(22) $$ And the primary
constraints are, $$ \eqalign {A^{ij}&:=\A =0\cr B^i&:=\B =0\cr
C^i&:=\C =0\cr D_{i\mu}&:=m_{i\mu}=0\cr \varphi^{i}&:=\pn =0\cr
\psi_1&:=\pu =0\cr \psi_2&:=\px =0.\cr}\eqno(23) $$

Apart from the constraints determining the auxiliary variables, the
moment $\pi$ is completely constrained. The Hamiltonian is not zero on
the primary constraints.

In the next section, we compute the secondary constraints.

\seccion Secondary constraints.

We compute the Poisson bracket of the Hamiltonian with all primary
constraints. From the conservation of $A^{ik}, B^i, C^i, D_{j\nu}$
one obtains the Lagrange multipliers, $$ \lambda_{kj} = -\pr
(n^ju')(n^ku') \eqno(24) $$ $$\eqalign {\eta_j &= -\pt 2(n^ju')
[((u'x')(ux')-x'^2(uu')]-\cr&- \omega^j\pt
[((ux'')x'^2-(ux')(x'x'')]\cr}\eqno(25) $$ $$\eqalign {\mu_j &= -\pt
2(n^ju') [((u'u')(ux')-u^2(x'u')]+\cr&+ \omega^j\pt
[((ux'')(ux')-u^2(x'x'')].\cr}\eqno(26) $$ If $\gamma^{i\nu}$ is
decomposed as $$ \gamma^{i\nu}=\alpha^iu^\nu +\beta^ix'^\nu
+\epsilon^{ik}n^{k\nu},\eqno(27) $$ then one obtains $$ \alpha^i
={\omega^ix'^2 - (n^iu') (ux')\over -g}\eqno(28) $$ $$ \beta^i
={-\omega^i(ux') + (n^iu') u^2\over -g}.\eqno(29) $$ The
antisymmetric part of $\epsilon^{ik}$ remains undetermined, while the
symmetric part is zero. $$ \epsilon^{\bar {ik}}={1\over 2}
(\epsilon^{ik} + \epsilon^{ki})=0\eqno(30) $$

The conservation of $\Psi_1, \Psi_2$ gives two secondary constraints,
$$ \Psi_3:= {\cal H}_0 = \ho =0\eqno(31) $$ $$ \Psi_4:= \hu
=0\eqno(32) $$ It shows that the Hamiltonian is zero.The conservation
of  $\varphi^i$ gives another secondary constraint, $$
{\eqalign{&\zeta^i:= -pn^i +{2\alpha\Phi\over \sqrt{-g}}(n^iu'') +
+{2\alpha\over \sqrt{-g}}\bigl [2\Phi '(n^iu')-\dot \Phi(n^ix'')\bigr
]+\cr &+ {2\alpha\Phi\over (-g)^{-3/ 2} }\bigl [ (x''n^i)\bigl
(u^2(x'u')-(ux')(uu')\bigr)- (u'n^i)\bigl ((ux')(ux'') -\cr&-u^2
(x'x'')-2(ux')(u'x')-2x'^2(uu')\bigr )\bigr ]=0.\cr}}\eqno(33) $$

The conservation of $\Psi_3$ and $\Psi_4$ is satisfied trivially. The
conservation of $\zeta^i$ gives an equation for $\omega^i$, $$
F^{ik}\omega^k + G^i=0,\eqno(34) $$ where $$ \eqalign { F^{ik}&=
\delta^{ik} {x'^2\over \rz} + \ppt\bigl [(uu')x'^2-(ux')(u'x')\bigr ]
-\delta^{ik}\pdt\bigl [(ux'')x'^2-\cr &-(ux')(x''x')\bigr ]-
\delta^{ik}(\pp)' -{2\alpha\over\sqrt{-g}}\bigl [(\partial_\mu \Phi
n^{k\mu})(x''n^i) + (\partial_\mu \Phi n^{i\mu})(x''n^k)\bigr
],\cr}\eqno(35) $$ and $$\eqalign {G^i&=
{u^2(x"n^i)-2(ux')(u'n^i)\over \sqrt{-g}}  +{2\alpha\over
\sqrt{-g}}(\partial_\mu\Phi n^{i\mu})(n^ku')^2 +\cr &+2\ppt
(n^iu')\bigl [u^2(x'u')-(ux')(uu')\bigr ]+\cr &  +2\pdt (n^iu')\bigl
[x'^2(uu')-(ux')(x'u')\bigr ] + 2{2\alpha\over \sqrt{-g}}
(\partial_\mu\Phi u'^\mu)(n^iu') +\cr &+  {2\alpha\over
\sqrt{-g}}\bigl [2(\partial_\mu\partial_\nu\Phi x'^\mu u^\nu) (n^iu')
-(\partial_\mu\partial_\nu\Phi u^\mu u^\nu) + (n^ix'')\bigr ].\cr}
\eqno(36) $$ The term independent of $\Phi$ in (35) is always
different from zero, provided $x'$ is a spatial vector. (34) is an
algebraic equation which allows for the computation of $\omega^i$
(for example, one can suppose analiticity in $\alpha$ ). It is
complicated, but we are not going to use it explicitly. The important
thing is that this Lagrange multiplier can be computed, and that the
conservation of $\zeta^i$ gives no other secondary constraint. This
is the complete set of constraints.

The constraints concerning the true variables of the theory can be
resumed in two covariant expressions that do not involve the
auxiliary variables. These expressions will be useful when fixing the
gauge. $$ \varphi_\mu:=\pi_\mu -\pr D_1x'_\mu=0\eqno(37) $$  is
equivalent to $\Psi_1=0$, $\Psi_2=0$,$\varphi^i=0$. Also, $$\eqalign
{\zeta_\mu&= p_\mu +
 {1\over\sqrt{-g}}\bigl [(ux')x'_\mu -x'^2u_\mu \bigr ] + \bigl
[{2\alpha(\partial_\nu \Phi u^\nu)\over \sqrt{-g}}+ 2\pt \tilde
\Gamma_{01}^1\bigr ] D_1x'_\mu -\cr &- \bigl [ 2\pt \tilde
\Gamma_{11}^1 +2\pp \bigr ] D_0x'_\mu - 2\pr (D_0x'_\mu)'- \pt \tilde
\Gamma_{01}^1 x''_\mu-\cr &- \pt \bigl [(ux')x'_\mu -x'^2u_\mu \bigr
](u'^\nu  D_0x'_\nu)+ \pr u"^\bot_\mu+\cr &+ \pt\bigl [u_\mu\bigl
((ux'')(x'u')-(ux')(u'x'')\bigr )+x'_\mu \bigl
(u^2(x''u')-(uu')(ux'')\bigr )\bigr ]\cr }\eqno(38) $$ is equivalent
to $\Psi_3 =0, \Psi_4 =0, \zeta^i =0$. We have used the  Christoffel
symbols of the metric $g_{ab}$, $\tilde \Gamma _{ab}^c = g\Gamma
_{ab}^c$. $$\eqalign{ \tilde \Gamma _{00}^0&=x'^2(uv)-(ux')(vx')\cr
  \tilde \Gamma _{00}^1&=u^2(vx')-(ux')(uv)\cr \tilde \Gamma
_{01}^0&=x'^2(uu')-(ux')(u'x') \cr
  \tilde \Gamma _{01}^1&=u^2(x'u')-(ux')(uu')\cr \tilde \Gamma
_{11}^0&=x'^2(ux'')-(ux')(x'x'')\cr
  \tilde \Gamma _{11}^1&=u^2(x'x'')-(ux')(ux'').\cr}\eqno(39) $$ The
covariant derivatives are, $$\eqalign {D_0x'^\mu&=u'^\mu -\Gamma
_{01}^0u^\mu- \Gamma _{01}^1x'^\mu\cr D_1x'^\mu&=x''^\mu -\Gamma
_{11}^0u^\mu- \Gamma _{11}^1x'^\mu .\cr}\eqno(40) $$ and $u''^\bot$
is the normal part to $u''$ given by $$ u''^{\bot\mu} =
u''^\mu-{1\over -g}[(u''x')(ux')-(u''u)x'^2]u^\mu-  {1\over
-g}[(u''u)(ux')-(u''x)u^2]x'^\mu .\eqno(41) $$ These are the
constraints one would have obtained if the original action, where the
auxiliary variables are substituted, had been used. We can see that
all the moments, $p$ and $\pi$ can be computed in terms of the
coordinates $x$ and $u$, so the degrees of freedom are notably
reduced. We expect some constraints to be first class in order to
contemplate the reparametrization invariance of the theory.

In the next section we study the character of these constraints and
compute the degrees of freedoom of the theory.

\seccion First and second class constraints.

Between the constraints associated to the auxiliary variables, there
is a first class constraint, corresponding to the undetermined
Lagrange multiplier. This constraint is
 $$ D_{\hat{ik}}= {1\over 2}(D_{i\mu}n_k^\mu -
D_{k\mu}n_i^\mu).\eqno(42) $$ The Poisson bracket of this constraint
with the remaining ones is zero. The gauge transformation it
generates only afects the variables $n^i$, and it is given by
 $$ \delta_\varepsilon n^{i\mu} = \varepsilon ^{\hat
{ik}}n^{k\mu}.\eqno(43) $$ The meaning of this transformation is an
infinitesimal rotation in the space of normal vectors. The other
constraints $A^{ij}$, $B^i$, $C^i$,$D_{\bar {ik}}$,  $D_{i\mu}u^\mu$,
and $D_{i\mu}x'^\mu$ are second class. All these constraints and the
gauge invariance (43) determine $n^{i\mu}$ and $m_{i\mu}$ without
ambiguity, so there is no dynamical variables. The remaining
constraints restrict the true variables of the theory. We are going
to elucidate which of them are first class.

The Hamiltonian we have computed, $$ {\cal H}= \Lambda_1\psi_1
+\Lambda_2\psi_2 + \tilde\psi_3\eqno(44) $$ with $$\eqalign{
 \tilde\Psi_3&=\Psi_3-\pr(n^iu')(n^ju')A_{ij}- (\pt 2(n^ju')
[(u'x')(ux')-x'^2(uu')]+\cr+& \omega^j\pt
[((ux'')x'^2-(ux')(x'x")])B_j-\cr-&
 (\pt 2(n^ju') [((u'u')(ux')-u^2(x'u')]-\cr-& \omega^j\pt
[((ux'')(ux')-u^2(x'x")])C_j +\cr+& {1\over
-g}[\omega^ix'^2-(n^iu')(ux')](D_{i\mu}u^\mu)+ {1\over -g}[-\omega
^i(x'u)-(n^iu')u^2](D_{i\mu}x'^\mu) +\cr
&+\omega^i\varphi^i,\cr}\eqno(45) $$ and $\omega^i$ given by equation
(34), is proportional to the first class constraints. $\Psi_1$ and
$\Psi_2$ are first class, as follows from direct computation. The
rest is a first class constraint we will call  $\tilde \Psi_3$. In
fact, one can show that $\Psi_3$ and $\Psi_4$ are not by themselves
first class constraints. In order to convert them into first class
constraints, they must be corrected with terms proportional to other
second class constraints. We will substitute $\Psi_3$ by $\tilde
\Psi_3$ in the set of constraints we are using to describe the
submanifold.

In order to obtain the remaining first class constraints, we make an
arbitrary linear combination of the  second class constraints $$
<F>=<a_{kl}A^{kl} +b_kB^k+c_kC^k
+d^{k\mu}D_{k\mu}+f_k\varphi^k+h_k\zeta^k+e\Psi_4>,\eqno(46)  $$ and
compute the Poisson bracket of this constraint with the others. This
procedure will determine the arbitrary coeficients until we have an
arbitrary linear combination of first class constraints. The result
is that they are all zero except
 $$ \eqalign{ &(d^in^j) =0 \cr &(d^ix') =-(n^ix'')e \cr
&(d^iu)=-(n^iu')e,\cr} \eqno(47) $$ and $e$ remains undetermined. The
only first class constraint is $$ \eqalign{\tilde\Psi_4&=\Psi_4+
+{1\over-g} [-(n^iu')x'^2+(n^ix'')(ux')](D_{i\nu}u^\nu)+\cr
&+{1\over-g} [-(n^iu')(ux')+(n^ix'')u^2](D_{i\nu}x'^\nu).
\cr}\eqno(48)$$

We have four first class constraints, so the degrees of freedoom of
the string in interaction with the dilaton are $D-2$, the same as the
free bosonic string. In the theory of Ref. [2] the interaction term
breaks the Weyl invariance, so it appears an additional degree of
freedoom. This problem is solved in this approach.

It is interesting to compare this result with the one for the rigid
string [4][5]. The rigid string is a theory which also presents two
dimensional reparametrization invariance, and presents second order
derivatives in the Lagrangean. The canonical formulation leads to
four first class constraints too, and the Poisson algebra of them
does not contain the Virasoro algebra as a subalgebra. This result is
achieved only when restricting to certain submanifold. For the rigid
string no second class constraints appear, so the degrees of freedoom
are twice the ones of the free string.

\seccion Light cone gauge.

We can impose four gauge fixing conditions. We take the light cone
gauge, defined by $$ \eqalign { \chi_1&=ux'\cr \chi_2&=u^2+ x'^2\cr
\chi_3&=x^+ -u_0^+\tau\cr \chi_4&=u^+ -u_0^+.\cr}\eqno(49) $$ When
$\Phi=0$, the constraints reduce to $$ p=u,\quad \pi=0 \eqno(50) $$
$\chi_1$ and $\chi_2$ are the constraints of the free bosonic string
while $\chi_3$ and $\chi_4$ correspond to the usual light cone gauge
conditions. In this formulation one has more degrees of freedoom and
we could select for $\chi_1$ and $\chi_2$ another conditions,
different from the usual constraints  $px'$ and $p^2 +x'^2$. If the
conditions were admissible, one would obtain an equivalent theory.

This gauge fixing allows the reduction to the physical modes, which
are the transversal modes $x^\top, u^\top$, computing the
longitudinal ones,  $$ \eqalign{ (ux')&=u^- x'^+ +u^+x'^- -(ux')^\top
\cr x'^-&=-(ux')^\top\over u^{+0}\cr u^-&=-x'^{\top 2}+u^{\top
2}\over 2 u^{+0}.\cr}\eqno(51) $$ We want to show now that (49) are
admissible gauge fixing conditions, by computing the Lagrange
multipliers. We take the total Hamiltonian, this is, proportional to
all first class constraints, $$ {\cal H}_T = \Lambda_1 \Psi_1
+\Lambda_2 \Psi_2 + \Lambda_3 \tilde\Psi_3 + \Lambda_4\tilde \Psi_4
\eqno(52)  $$ and compute the conservation of all gauge fixing
conditions. The result is $$ \eqalign{ \Lambda_3&-1\cr \Lambda_2 &=
-{u^{+0}u'^- -(uu')^\top\over x'^{\top 2}}\cr \Lambda_1 &=
-{\Lambda'_4 x'^{\top 2}+(u'x')^\top\over u^{+0}u^- +u^{\top
2}}\cr}\eqno(53) $$ and $\Lambda_4$ satisfies the equation $$
u^{+0}\Lambda'_4=-\omega^in^{i+}-u^{+0}{ (u'x')^\top\over x'^{\top
2}}.\eqno(54) $$

Regretfully, the light cone gauge does not leave the action in
canonical form  $$ S_{phys}=\int d^2\xi (-(p\dot x)^\top -(\pi \dot
u)^\top +p^+\dot x^- + \pi^+\dot u^-),\eqno(55) $$ so the computation
of the Hamiltonian is not direct from here. Nevertheless, the
equations of motion for the transversal modes $x^{\top}$, $u^{\top}$
are first order (in time) equations. We can compute the energy of the
system as the conserved quantity associated to traslational
invariance of $S_{phys}$. It only holds (in this gauge) if
$\partial_+\Phi=0$. This means that the light cone gauge is
apropriate to compute the energy only in this case. $$ E=\int
d\sigma\bigr [{ \delta{\cal L}\over\delta \dot x^{\top}}\dot x^{\top}
+ { \delta{\cal L}\over \delta\dot u^{\top}}\dot u^{\top}-{\cal
L}\bigl ],\eqno(56)  $$  where $$ S_{phys}=\int d\sigma d\tau {\cal
L}. \eqno(57) $$ The result is, $$ E=\int d\sigma u^{+0}p^{-}\eqno(58)
$$ whith $p^{-}$ expressed in terms of the physical modes $$ \eqalign{
&p^{-}= {1\over 2 u^{+0}}(x'^{\top 2} +u^{\top 2}) +2\alpha \Phi
x'^{\top 2}\cdot \cr &\bigl [(\partial _\mu\Phi x'^\mu)\bigl
(-{(x'x'')^\top\over 2 u^{+0}}  -{(uu')^\top\over  u^{+0}} +{x'^{\top
2}(x'u')^\top(x'u)^\top\over  u^{+0}} +{x'^{\top 2}(x'x'')^\top
u^{\top 2}\over  2u^{+0}}\bigr )- \cr &-(\partial _\mu\Phi
u^\mu)\bigl (-{(x'u')^\top\over 2 u^{+0}}  -{(ux'')^\top\over
u^{+0}} +{x'^{\top 2}(x'x'')^\top(x'u)^\top\over  u^{+0}} +{x'^{\top
2}(x'u')^\top u^{\top 2}\over  2u^{+0}}\bigr )\bigr ] \cr }\eqno(59)
$$

The Hamiltonian only depends on  $\Phi$ through his derivatives, so
when  $\Phi=$ctt one recovers the energy of the free bosonic string.
If $\Phi\not=$ctt, in the energy appears a term proportional to
$\alpha$. This term produces the energy not being definite in sign.

We compare with the system treated in  [2]. This system has a degree
of freedoom more. The canonical coordinates one uses to describe it
are $q^s=(x^\mu,\gamma)$, $p_s=(p_\mu,\pi)$, and the action is
 $$ S=\int d\sigma d\tau\bigl[p\dot
x+\pi\dot\gamma-\lambda_0T_0-\lambda_1T_1\bigr] \eqno(60) $$ where
$T_0$ and $T_1$ are the first class constraints which satisfy the
Virasoro algebra. They are given by the following expressions $$
T_0={1\over 2}G^{rs}p_rp_s+{1\over
2}G_{rs}q'^rq'^s-2(N_sq'^s)'\eqno(61) $$
 and $$ T_1=p_sq^s-2(N^sp_s)',\eqno(62) $$ where we have used the
notation $N_s=(-\partial_\mu\phi,0)$, $Ns=(0,1)$ and $$
G_{rs}=\left(\matrix{\eta_{\mu\nu}&-\partial_\mu\phi\cr
-\partial_\nu\phi&0}\right).\eqno(63) $$

If we fix the light cone gauge,  $$ x^+=p^{+0}\tau,$$
$$p^+=p^{+0},\eqno(64) $$ the physical Hamiltonian is $p^-$, which in
terms of the physical modes is $$\eqalign{
&p^-\left[p^{+0}-{\partial^+\phi^+\pi\over(\partial\phi)^2}\right]
=\cr &{1\over2}(p^{\top 2} +x'^{\top 2}) + {\pi^2\over
2(\partial\phi)^2}+ {p^{+0}\partial^-\phi\pi\over(\partial\phi)^2}+
{(\partial\phi p)^\top\pi\over(\partial\phi)^2}
+{\phi'\gamma'\over(\partial\phi)^2}-2\phi'',\cr}\eqno(65) $$ and it
is not positive definite.

Our conclussion is that the interaction term of the string and the
dilaton must be corrected in order to obtain a consistent quantum
theory.

\vfill \eject

\seccion Conclusions.

In this paper we obtained the canonical formulation of the string in
interaction with a background dilaton field, with an action of
Nambu-Goto type which has second order time derivatives. The complete
set of constraints is computed, and it is found that four first class
constraints appear, reflecting the reparametrization invariance of
the lagrangean action. In addition, the theory is restricted by
second class constraints. We decouple covariantly the first and
second class constraints. However, because of the second class
constraints, the covariant quantization of the system becomes
intrincated. The degrees of freedoom of the theory are the same as
for the free bosonic string, in distinction to the  Polyakov type
theory, which has only first order time derivatives. We had used the
light cone gauge to reduce to the physical modes, and to compute the
physical hamiltonian, which becomes indefinite in sign if the dilaton
field is different from constant. It is well known that higher order
terms in the curvature should be included in order to obtain the low
energy approximation of a complete string theory. It is the
Hamiltonian of the complete theory the one which is required to be
positive definite. Our result, which clearly extend to the case when
the other background fields are not trivial, shows that any
conclusion based on an analysis of a truncated theory could be
modified by higher order contributions. The theory is compared with
other approaches. \vfill \eject

\centerline {\bf REFERENCES} \vskip 1cm \item{[1]} C. Lovelace, {\sl
Phys. Lett.} {\bf B135} (1984), {\sl Nucl. Phys.}{\bf B273}(1986) 413.
\item{}E. Fradkin and A.A. Tsuytlin, {\sl Nucl. Phys.} {\bf B261}
(1985) 1. \item{} C. G. Callan, D. Friedan, E.J. Martinec and M. J.
Perry, {\sl Nucl. Phys.} {\bf B262} (1985) 593. \item{} J. Maharana
and G. Veneziano, {\sl Nucl. Phys.} {\bf B283} (1987) 126. \item{} R.
Akhoury and Y. Okada, {\sl Phys. Rev.} {\bf D34} (1986) 3760.
\item{} H. J. de Vega and N. S\'anchez, {\sl Nucl. Phys.} {\bf B299}
(1988) 818. \item{} G. T. Horowitz and A. R. Steif, {\sl Phys. Rev.}
{\bf D42} (1990) 1950. \item{[2]}I.L. Buchbinder, E.S. Fradkin,S.L.
Lyakhovich, V. D. Pershin, {\sl  Int. J. Mod. Phys.} {\bf 3} (1991)
\item{[3]} M.A. Lled\'o, A. Restuccia, "Canonical Formalism,  the
Initial Value Problem and the Maximum Principle" {\sl   Relativity
and Gravitation: Classical and Quantum.  Proceedings of Silarg VII.}
World Scientific Publishing.  (1991). 227-232.
 \item{[4]} M. A. Lled\'o and A. Restuccia, {\sl  Ann. Phys.} {\bf
224} (1993). \item{}M. A. Lled\'o, A. Restuccia, "Theories With Higher
 Order Derivatives: The Rigid String" {\sl   Proceedings XIX
International Colloquium on Group Theoretical Methods in  Physics.}
Anales de F\' {\i}sica, Monograf\' {\i}as. M.O., M.S. and J.M.G.
 (Eds). CIEMAT/RSEF, Madrid (1993)126. \item{[5]} A.Polyakov, {\sl
Nucl. Phys.} {\bf B268} (1986) 406. \end